\documentclass[showpacs,preprintnumbers,amsmath,amssymb,twocolumn]{revtex4}

\usepackage{graphicx}
\usepackage{dcolumn}
\usepackage{bm}
\usepackage{color}

\begin{document}


\title{Energy Extraction and Particle Acceleration Around Rotating Black Hole in Ho\v{r}ava-Lifshitz Gravity}

\author{Ahmadjon Abdujabbarov}%
   \email{ahmadjon@astrin.uz}

\author{Bobomurat Ahmedov}
 \email{ahmedov@astrin.uz}

 \affiliation{%
Institute of Nuclear Physics, Ulughbek, Tashkent
              100214, Uzbekistan\\
    Ulugh Beg Astronomical Institute,    Astronomicheskaya 33,
    Tashkent 100052, Uzbekistan\\
    Inter University Centre for Astronomy \& Astrophysics,
    Post Bag 4, Pune 411007,
    India}

\author{Bahodir Ahmedov}
 \email{bahat61@mail.ru}

\affiliation{Applied Mathematics and Informatics Department,
National University of Uzbekistan, Tashkent 100095, Uzbekistan}
\date{\today}

\begin{abstract}

Penrose process on rotational energy extraction of the black hole
(BH) in the original non-projectable Ho\v{r}ava-Lifshitz gravity
is studied. The strong dependence of the extracted energy from the
special range of parameters of the Ho\v{r}ava-–Lifshitz gravity,
such as parameter $\Lambda_W$ and specific angular momentum $a$
has been found. Particle acceleration near the rotating BH in
Ho\v{r}ava-–Lifshitz gravity has been studied. It is shown that
the fundamental parameter of the Ho\v{r}ava-–Lifshitz gravity can
impose limitation on the the energy of the accelerating particles
preventing them from the infinite value .

\end{abstract}

\maketitle

\section{\label{sec:intro}Introduction}

Recently Ho\v{r}ava proposed a UV (Ultra-Violet) complete,
non-relativistic gravity theory which is power-counting
renormalizable one giving up the Lorentz invariance~\cite{h1,h2}.
Since then, many authors paid attention to this scenario to apply
it to the black hole (BH)
physics~\cite{ghodsi,CaiCao,Konoplya22,SChen,CastilloLar,ChenYang},
cosmology~\cite{Takashi,CalcagniJHEP,Kalyana,Wang,loboooo,mukohayama,Saridakis}
and observational tests~\cite{lobo1}. Here we investigate the
Penrose process around rotating BHs in the Ho\v{r}ava–-Lifshitz
gravity theory. The quantum interference effects~\cite{htaa} and
the motion of the test particle around BH~\cite{aha} in
Ho\v{r}ava–-Lifshitz gravity have been also recently studied.

In the paper \cite{lobo1} the possibility of observationally
testing Ho\v{r}ava–-Lifshitz gravity at the scale of the Solar
System, by considering the classical tests of general relativity
(perihelion precession of the planet Mercury, deflection of light
by the Sun and the radar echo delay) for the Kehagias-Sfetsos (KS)
asymptotically flat black hole solution of Ho\v{r}ava–-Lifshitz
gravity has been considered. Recently authors of the
paper~\cite{claus} have studied the particle motion in the
space-time of a KS black hole. The stability of the Einstein
static universe by considering linear homogeneous perturbations in
the context of an Infra-Red (IR) modification of
Ho\v{r}ava–-Lifshitz gravity has been studied in~\cite{lobo2}.
Potentially observable properties of black holes in the deformed
Ho\v{r}ava–-Lifshitz gravity with Minkowski vacuum: the
gravitational lensing and quasinormal modes have been studied
in~\cite{konoplya11}. {The authors of the paper \cite{einstein}
derived the full set of equations of motion, and then obtained
spherically symmetric solutions for UV completed theory of
Einstein proposed by Ho\v{r}ava. }

{Black hole solutions and the full spectrum of spherically
symmetric solutions in the five-dimensional nonprojectable
Ho\v{r}ava–-Lifshitz type gravity theories have been recently
studied in~\cite{tsoukalas1}. Geodesic stability and the spectrum
of entropy/area for black hole in Ho\v{r}ava–-Lifshitz gravity via
quasi-normal modes approach are analyzed in~\cite{dav1}. Particle
geodesics around Kehagias-Sfetsos black hole in
Ho\v{r}ava–-Lifshitz gravity are also investigated by authors of
the paper~\cite{gwak}. Recently observational constraints on
Ho\v{r}ava–-Lifshitz gravity have been found from the cosmological
data~\cite{saridakis3}. Authors of the paper~\cite{ghodsi} have
found all spherical black hole solutions for two, four and six
derivative terms in the presence of Cotton tensor.}

Recently the rotating black hole solution in the context of the
Ho\v{r}ava–-Lifshitz  gravity has been obtained in \cite{ghodsi}.
In this paper we plan to study the energy extraction mechanism
through the Penrose process and particle acceleration mechanisms
near the rotating black hole in the Ho\v{r}ava–-Lifshitz gravity.
Authors of the paper \cite{banados} considered Kerr black hole as
particle accelerators to arbitrary high energies. The results of
the paper~\cite{banados} have been commented in \cite{berti} where
authors concluded that astrophysical limitations on the maximal
spin, back-reaction effects and sensitivity to the initial
conditions impose severe limits on the likelihood of such
accelerations. New proposed solution forces us to study the
particle acceleration in the gravity theory of
Ho\v{r}ava–-Lifshitz.  Recently  Patil and  Joshi \cite{joshi1}
have shown that the naked singularities that form due to the
gravitational collapse of massive stars provide a suitable
environment where particles could get accelerated and collide at
arbitrarily high center-of-mass energies.

{The paper is organized as follows.} The description of the
rotating black hole solution and ergosphere around it considered
in the Sec.~\ref{metriccc}. Penrose process in the ergosphere of
the rotating black hole in Ho\v{r}ava--Lifshitz gravity has been
studied in Sec.~\ref{enextract}. Sec.~\ref{accelerate} is devoted
to study the particle acceleration mechanism near the black hole
in Ho\v{r}ava–-Lifshitz gravity. We conclude our results in
Sec.~\ref{conclusion}.

We use a system of units in which $c = G = 1$, a space-like
signature $(-,+,+,+)$ and a spherical coordinate system
$(t,r,\theta ,\varphi)$. Greek indices are taken to run from 0 to
3, Latin indices from 1 to 3.

\section{\label{metriccc} Extreme Rotating Black Hole in Ho\v{r}ava--Lifshitz Gravity}

The four-dimensional metric of the spherical-symmetric spacetime
written in the ADM
formalism~\cite{lobo1,loboetal,lobo2,konoplya11} has the following
form:
\begin{equation}\label{metri}
ds^{2}=-N^{2}c^{2}dt^{2}+g_{ij}(dx^{i}+N^{i}dt)(dx^{j}+N^{j}dt)\ ,
\end{equation}
where $N$ is the lapse function, $N^i$ is the shift vector to be
defined.

The Ho\v{r}ava–-Lifshitz action describes a nonrelativistic
renormalizable theory of gravitation and is given by (see for more
details Ref.~\cite{h1,h2,lobo1,loboetal,lobo2,konoplya11})
\begin{eqnarray}\label{action}
& \hspace{-0.5cm}S =& \int dtdx^3 \sqrt{-g} N
\bigg[\frac{2}{\kappa ^2}(K_{ij}K^{ij}-\lambda_g K^2)\nonumber\\
&&+\frac{ \kappa^2\mu}{2\nu_g^2}\epsilon^{ijk}R_{il}\nabla_j
R^l_{\
k} -\frac{\kappa^2 \mu^2}{8}R_{ij}R^{ij}+\frac{\kappa^2\mu^2}{8(3\lambda_g-1)} \nonumber \\
&&\times \left(\frac{4\lambda_g-
1}{4}R^2-\Lambda_WR+3\Lambda_W^2\right)-\frac{\kappa^2}{2\nu_g^4}C_{ij}C^{ij}
\bigg] ,
\end{eqnarray}
where $\kappa, \lambda_g, \nu_g, \mu$,  and $\Lambda_W$ are
constant parameters, the Cotton tensor is defined as
\begin{equation}
C^{ij}=\epsilon^{ikl}\nabla_k (R^{j}_{\ l}-\frac{1}{4} R\delta^j_l
)\ ,
\end{equation}
$R_{ijkl}$ is the three-dimensional curvature tensor, and the
extrinsic curvature $K_{ij}$ is defined as
\begin{equation}
K_{ij}=\frac{1}{2N}(\dot{g}_{ij}-\nabla_i N_j -\nabla_j N_i)\ ,
\end{equation}
where dot denotes a derivative with respect to coordinate $t$.

If one considers up to second derivative terms in the action
(\ref{action}), one can find the known topological rotating
solutions given by~\cite{ghodsi4} for equations of motion in the
Ho\v{r}ava-–Lifshitz gravity. Since we are considering matter
coupled with the metric in a relativistic way, we can consider the
metric in Boyer-Lindquist coordinates  instead of its ADM form
which is the solutions of Ho\v{r}ava-Lifshitz gravity.  In the
Einstein's gravity this spacetime metric reads in Boyer-Lindquist
type coordinates in the following form (see, e.g.~\cite{ghodsi}):
\begin{eqnarray}\label{metric}
& ds^2=& -\frac{\Delta_{\rm r}}{\Sigma^2 \rho^2} \left[dt-a \sin^2
\theta d\varphi \right]^2+\frac{\rho^2}{\Delta_{\rm r}} dr^2
\nonumber\\&& + \frac{\rho^2}{\Delta_{\rm \theta}}d\theta^2 +
\frac{\Delta_{\rm \theta}\sin^2 \theta}{\Sigma^2
\rho^2}\left[adt-(r^2+a^2)d\varphi\right]^2 ,\ \
\end{eqnarray}
where the following notations
\begin{eqnarray}
&& \Delta_{\rm r}= (r^2 + a^2)\left(1+\frac{r^2}{l^2}\right)-2Mr\
, \nonumber\\ && \Delta_{\rm
\theta} = 1-\frac{a^2}{l^2}\cos^2 \theta \ , \nonumber\\
&& \rho^2= r^2+a^2\cos^2\theta\ ,  \nonumber\\ &&
\Sigma=1-\frac{a^2}{l^2}\ , l^2=-2/\Lambda_W \nonumber
\end{eqnarray}
are introduced, $M$ is the total mass of the central BH, $a$ is
the specific angular momentum of the BH. Note that metric
(\ref{metric}) in ADM form can be written as~\cite{ghodsi4}:
\begin{eqnarray}\label{metricadm}
& ds^2=& -\frac{\rho^2 \Delta_{\rm r} \Delta_{\rm
\theta}}{\Sigma^2 \Xi^2} dt^2 +\frac{\rho^2}{\Delta_{\rm r}} dr^2
 + \frac{\rho^2}{\Delta_{\rm \theta}}d\theta^2 \nonumber\\&& +
\frac{\Xi^2 \sin^2 \theta}{\Sigma^2 \rho^2}\left[d\varphi -\varpi
dt \right]^2 ,\ \
\end{eqnarray}
where
\begin{eqnarray}
&& \Xi^2= (r^2 + a^2)^2 \Delta_{\rm \theta} -a^2 \Delta_{\rm r}
\sin^2\theta \ , \nonumber\\
&& \varpi=-\frac{a}{\Xi^2} [ -(r^2+a^2)\Delta_{\rm \theta}
+\Delta_{\rm r}] \nonumber
\end{eqnarray}

The spacetime (\ref{metric}) has a horizon where the four velocity
of a corotating observer tends to zero, or the surface $r= const $
becomes null. Thus we have:
\begin{eqnarray}
r_{+}\simeq (1-3\delta)[M+\sqrt{M^2-a^2(1+3 \delta)}] ,
\end{eqnarray}
where we have introduced small dimensionless parameter $\delta=
a^2/l^2\ll 1$.

The static limit is defined where the time-translation Killing
vector $\xi_{(t)}^\alpha$ becomes null (i.e. $g_{00}=0$), so the
static limit of the BH can be described as
\begin{eqnarray}
r_{\rm st}& \simeq & (1-3\delta)\bigg\{M \nonumber\\&&
\hspace{-0.75 cm}+\sqrt{M^2-a^2(1+\delta\sin^2\theta)(1+3
\delta)\cos^2\theta }\bigg\}  .\
 \end{eqnarray}
In the recent paper~\cite{ghodsi4} authors provide the ADM and
Boyer-Lindquist forms of the spacetime metric in
Ho\v{r}ava-Lifshitz gravity both. From the expression
(\ref{metricadm}) one can easily see that the results for the
radii of event horizon and static limit will be identical in the
the ADM and Boyer-Lindquist ones (For more details about these
forms of spacetime metric presentation in Ho\v{r}ava-Lifshitz
gravity we refer  to the papers~\cite{ghodsi}
and~\cite{ghodsi4}.).

Considering only the outer horizon, $r_{+}$ and static limit, $
r_{\rm st} $, it can be verified that the static limit always lies
outside the horizon. The region between the two is called the
ergosphere, where timelike geodesics cannot remain static but can
remain stationary due to corotation  with the BH with the specific
frame dragging velocity at the given location in the ergosphere.
This is the region of spacetime where timelike particles with
negative angular momentum relative to the BH can have negative
energy relative to the infinity. Then, energy could be extracted
from the hole by the well-known Penrose process~\cite{penrose69}.

\begin{figure*}
a) \includegraphics[width=0.45\textwidth]{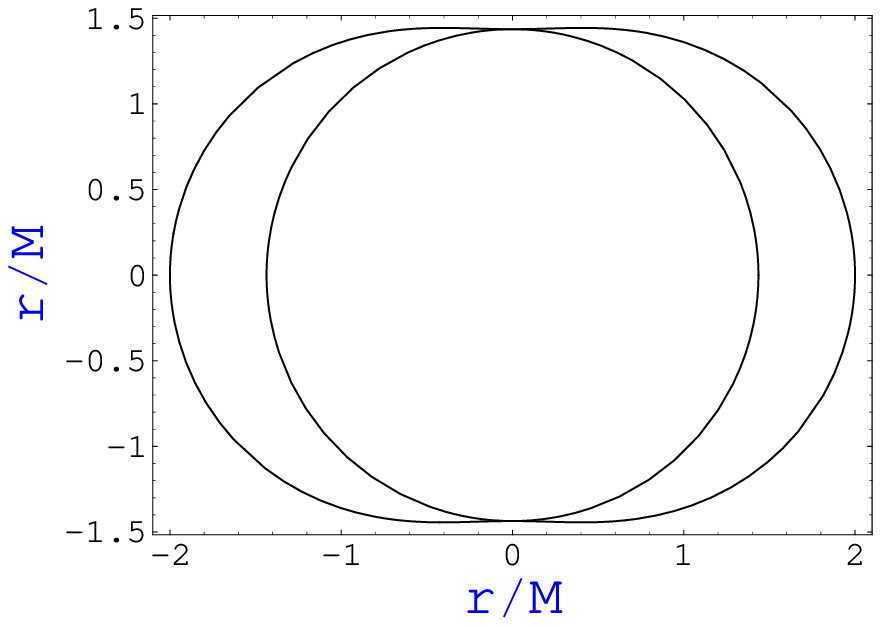}
b) \includegraphics[width=0.45\textwidth]{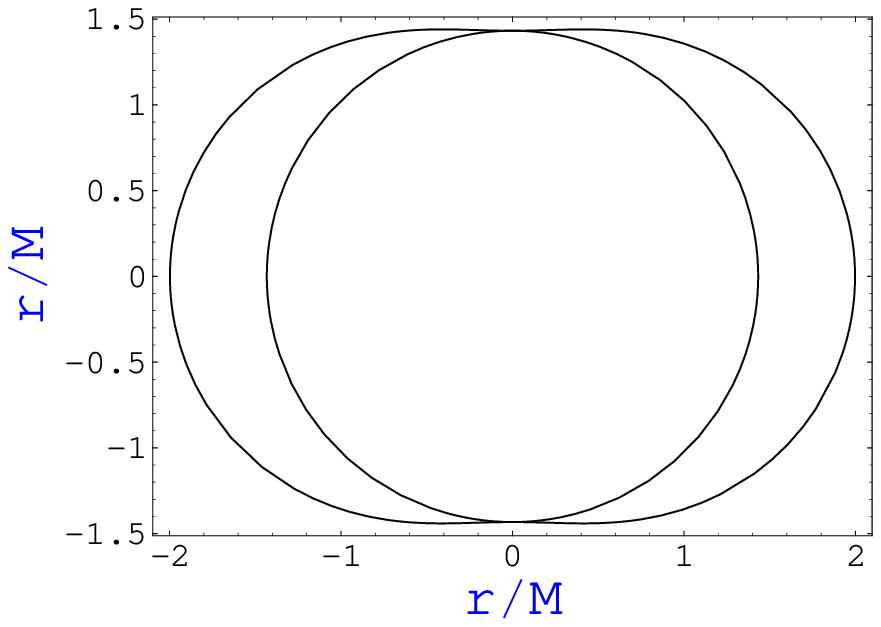}

c) \includegraphics[width=0.45\textwidth]{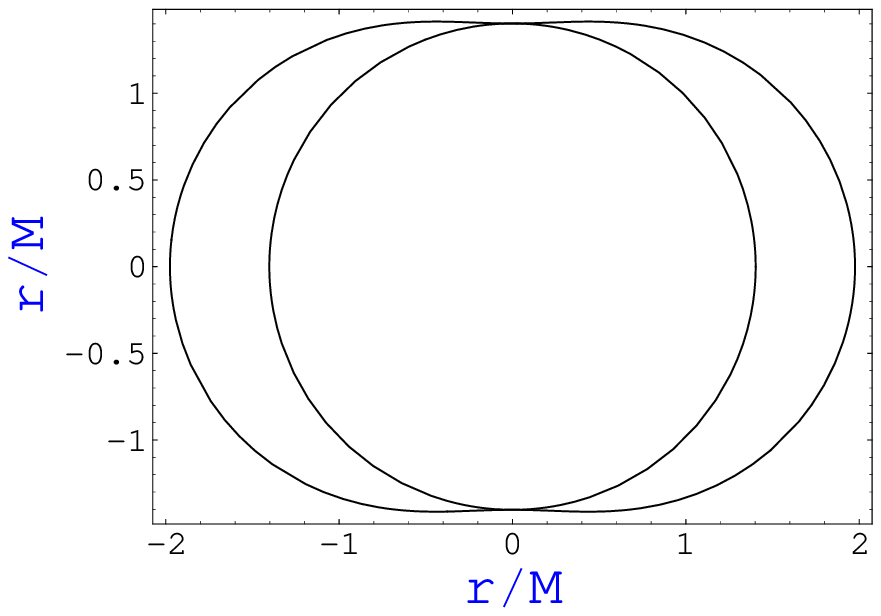} d)
\includegraphics[width=0.45\textwidth]{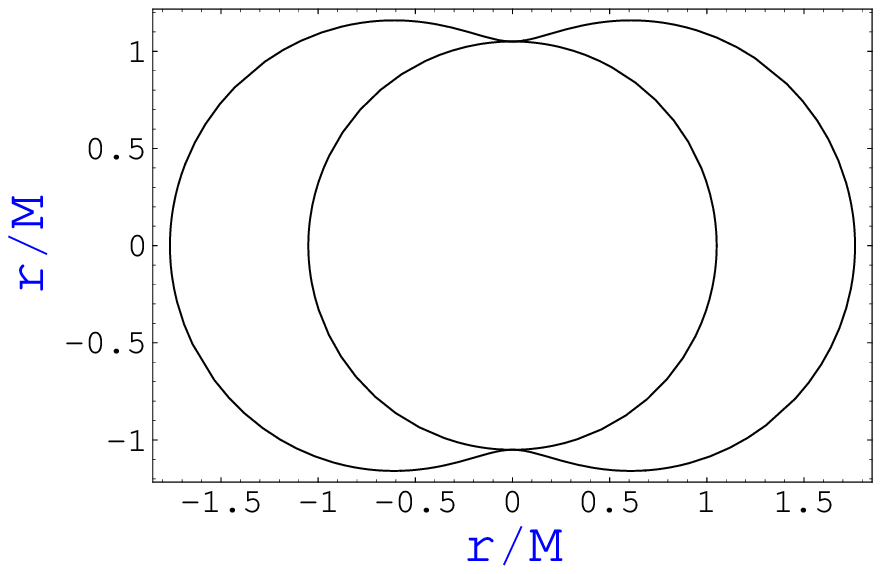}
\caption{\label{ergosphere} The dependence of the shape of the
ergosphere from the small dimensionless parameter $\delta$: a)
$\delta=0$, b) $\delta=0.001$ c) $\delta=0.01$, d) $\delta=0.1$.}
\end{figure*}

In Fig. \ref{ergosphere} the dependence of the shape of the
ergosphere from the small dimensionless parameter $\delta$ is
shown. From the figure one can see, that the relative shape of the
ergosphere becomes  bigger with  increasing the module of the
parameter $\delta$. Although  in the polar region there is no
ergoregion in the presence of the nonvanishing $\delta$ parameter,
where Penrose process can be realized, near to polar zone
ergoregion becomes more bigger than that in the Kerr spacetime. It
may increase the efficiency of the Penrose process.

\section{\label{enextract}Energy Extraction of Black Hole Through Penrose Process}

Due to the existence of an ergosphere around the BH, it is
possible to extract energy from BH by means of the Penrose
process. Inside the ergosphere, it is possible to have a timelike
or null trajectory with the negative total energy. As a result,
one can envision a particle falling from infinity into the
ergosphere and splitting into two fragments, one of which attains
negative energy relative to the infinity and falls into the hole
at the pole, while the other fragment would come out by
conservation of energy with the energy being greater than that of
the original incident particle. This is how the energy could be
extracted from the hole by axial accretion of particles with the
nonvanishing  momentum and $\delta$ parameter.

Consider the equation of motion of such negative energy particle
at the equatorial plane ($\theta=\pi/2\ ,\ \dot{\theta}=0$).
%
Using the Hamilton-Jacobi formalism the energy $E$ and angular
momentum $L$ of the particle are given as (see, e.g~\cite{mtw})
\begin{eqnarray}
&-\tilde{E}=-\frac{E}{m}=&\left[-\frac{1}{\Sigma^2}\left(1-\frac{2M}{r}-
\frac{r^2+a^2}{l^2}\right)\right]\dot{t}\nonumber\\
&& +
 \frac{a}{\Sigma^2}\left(\frac{r^2+a^2}{l^2}
-\frac{2M}{r}\right)
\dot{\varphi}\label{energy1} \ ,\\
& \tilde{L}=\frac{L}{m}= & \frac{1}{\Sigma^2} \left(
r^2+a^2\frac{l^2-r^2-a^2}{l^2}+
\frac{2Ma^2}{r} \right) \dot{\varphi} \nonumber\\
&& +
 \frac{a}{\Sigma^2}\left(\frac{r^2+a^2}{l^2}
-\frac{2M}{r}\right) \dot{t}\label{momentum1} \ ,
\end{eqnarray}

From the equations (\ref{energy1}) and (\ref{momentum1}) one can
easily obtain the equation of motion as:
\begin{eqnarray}\label{hamjameqq}
\alpha E^2 + \beta E +\gamma +\frac{\rho^2}{\Delta_{\rm r}}
\dot{p^{\rm r}}^2+ m^2 = 0\ ,
\end{eqnarray}
where we have introduced  the following notations:
\begin{eqnarray}
& \alpha= & \frac{1}{\Sigma^2}\left(r^2+a^2-a^2
\frac{r^2+a^2}{l^2}+a^2\frac{2M}{r}\right)
\Gamma^{-1} \ , \\
&\beta= & \frac{2aL}{\Sigma^2}
\left(\frac{r^2+a^2}{l^2}-\frac{2M}{r}\right) \Gamma^{-1}\ , \\
& \gamma = & -\frac{L^2}{\Sigma^2}
\left(1-\frac{2M}{r}+\frac{r^2+a^2}{l^2}\right) \Gamma^{-1}\ , \\
\end{eqnarray}
and
\begin{eqnarray} & \Gamma =&
-\frac{1}{\Sigma^4}\left(1-\frac{2M}{r}+
\frac{r^2+a^2}{l^2}\right)
\nonumber\\
&& \times \left(r^2+a^2 \frac{l^2-a^2-r^2}{l^2}+ a^2 \frac{2M}{r}
\right)
\nonumber\\
&& -\frac{a^2}{\Sigma^4} \left(\frac{r^2+a^2}{l^2}-\frac{2M}{r}
\right)^2\ .
\end{eqnarray}

From the equations (\ref{energy1}), (\ref{momentum1})
(\ref{hamjameqq}) one can easily obtain the equations of motion in
the following form:
\begin{eqnarray}
\frac{dt}{ds}&=&\frac{\Sigma^2}{r^2 \Delta_{\rm r}}\Big\{
[(r^2+a^2)^2-\Delta_{\rm r}
a^2] E \nonumber\\
&&  + (\Delta_{\rm r}-r^2-a^2)a L\Big\}\ ,\label{tuch}\\
\frac{d\varphi}{ds} &= & \frac{\Sigma^2}{r^2 \Delta_{\rm r}}
\Big\{(\Delta_{\rm r}-a^2) L +(r^2+a^2-\Delta_{\rm r})E\Big\}\ , \
\ \ \ \label{phiuch}
\\
\left(\frac{dr}{ds}\right)^2&=& E^2-V_{\rm eff}\ ,  \label{ruch}\\
V_{\rm eff}&=&\left(1+\frac{\Delta_{\rm r} \alpha}{\rho^2}\right)
E^2+ \frac{\Delta_{\rm r}}{\rho^2} \left(\beta E +\gamma
+1\right)\nonumber .
\end{eqnarray}

\begin{figure}
\includegraphics[width=0.45\textwidth]{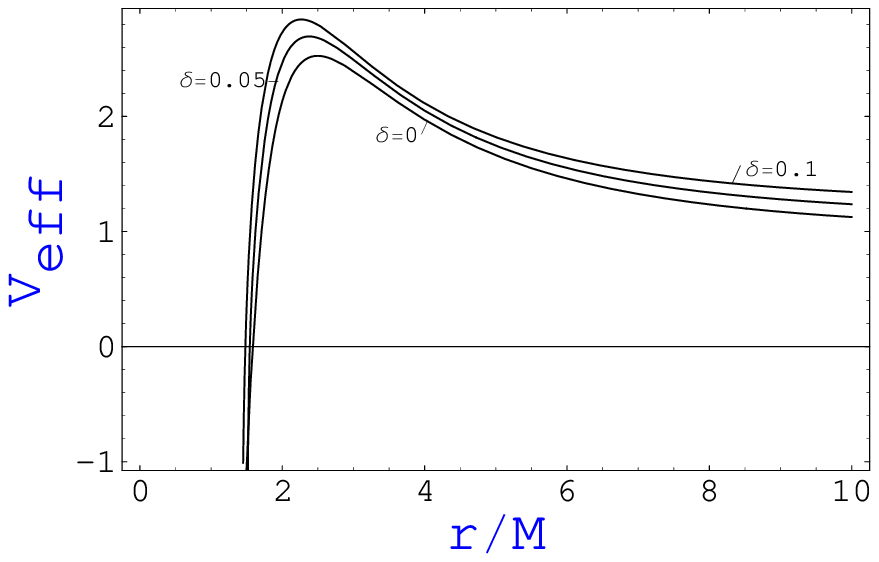}

\caption{\label{effpot} The radial dependence of the effective
potential of  radial motion of the massive test particle for the
different values of the dimensionless parameter $\delta$.}
\end{figure}

In the Fig. \ref{effpot} the radial dependence of the effective
potential of  radial motion of the massive test particle has been
shown for the different values of the parameter $\delta$. Here for
the energy and momenta of the particle the following values are
taken: $E/m=0.9$, $L/mM=6.2$. The presence of the parameter
$\delta$ slightly shifts the shape of the effective potential up.

When the one of two produced particles falls into the central BH,
the mass of the BH will change by $\Delta M = E$. The change in
mass can be made as large as one pleases by increasing the mass
$m$ of the infalling particle. However, there is a lower limit on
$\Delta M$ which could be added to the BH corresponding to $m= 0$
and $ \dot{p^{\rm r}} = 0$~\cite{mtw}. Evaluating all of the
required quantities at the horizon $r_{+}$, one can  easily get
the limit for the change in BH mass as
\begin{eqnarray}\label{minenergy}
E_{\rm min}=L \frac{\delta
(a^2+r_{+}^2)/a-{2Ma}/{r_{+}}}{r_{+}^2+a^2- a^2\delta-
r^2\delta+a^2{2M}/{r_{+}}}\ .
\end{eqnarray}

From the expression (\ref{minenergy}) one may conclude that
Penrose process can be realized if the condition
$\delta<{2Ma^2}/{r_{+} (a^2+r_{+}^2)}$ will be satisfied. Since
current astrophysical data indicate that parameter $\delta$ is
much less than 1, one may conclude that Penrose process is more
realistic process among the energy extraction mechanisms  from BH
in Ho\v{r}ava–-Lifshitz scenario. However, it should be mentioned
that in the early Universe when the module of the cosmological
constant played important role, energy extraction from the
rotating BH could be impossible in Ho\v{r}ava--Lifshitz scenario
due to the positivity of the sign of $E_{\rm min}$. This
limitation for Penrose process does not exist in the standard
theory gravity and appears in the modified theory gravity as
Ho\v{r}ava--Lifshitz one.


\section{\label{accelerate} Particle Acceleration Near the Black Hole}

Let us find the energy $E_{\rm cm}$ in the center of mass of
system of two colliding particles with energy  at infinity $E_{1}$
and $E_{2}$ in the gravitational field described by spacetime
metric (\ref{metric}). It can be obtained from
\begin{equation}
\left(\frac{1}{\sqrt{-g_{00}}} \ E_{\rm cm}, 0, 0, 0\right)=
m_{1}v_{(1)}^{\alpha}+m_{2}v_{(2)}^{\alpha},
\end{equation}
where $v_{(1)}^{\mu}$ and $v_{(2)}^{\nu}$ are the 4-velocities of
the particles, properly normalized by
$g_{\mu\nu}v^{\mu}v^{\nu}=-1$ and $m_{1}$, $m_{2}$ are rest masses
of the particles. We will consider two  particles with equal mass
($m_{1}=m_{2}=m_{0}$) which has the energy at infinity
$E_{1}=E_{2}\simeq 1$. Thus we have
\begin{eqnarray}
E_{\rm cm}= m_{0} \sqrt{2} \sqrt{1- g_{\alpha\beta}
v_{(1)}^{\alpha} v_{(2)}^{\beta}}.
\end{eqnarray}

Now using the equations (\ref{tuch})-(\ref{ruch}) one can obtain
expression for the energy of colliding particles near the
Ho\v{r}ava–-Lifshitz black hole as:

\begin{widetext}
\begin{eqnarray}
E_{\rm c.m.}^2 & = & \frac{2 m_0^2 \Sigma^2}{r
\Delta_r}\Bigg\{\frac{\delta r^5}{
a^{2}}+2r^2[r(1+2\delta)-1-4\delta]-[2a-\delta
ra(1+r^2)](l_1+l_2)+l_{1}l_{2}[2-r+\delta r (1+\frac{r^2}{ a^{2}})] \nonumber\\
&& + \,  2 a^2[1+(1+1.5 \delta) r]-\sqrt{2(a-l_{1})^2-(a^2\delta -
2a\delta l_{1}+l_{1}^2 +l_{1}^2 \delta+2r)r -2\delta +2
\frac{\delta l_{1}}{a}-\frac{l_{1}^2+r^2}{a^2}\delta
r^3}\nonumber\\
&&\times \sqrt{2(a-l_{2})^2-(a^2\delta - 2a\delta l_{2}+l_{2}^2
+l_{2}^2 \delta+2r)r -2\delta +2 \frac{\delta
l_{2}}{a}-\frac{l_{2}^2+r^2}{a^2}\delta r^3}\, \Bigg\}
\end{eqnarray}
\end{widetext}

\begin{figure}
\includegraphics[width=0.5\textwidth]{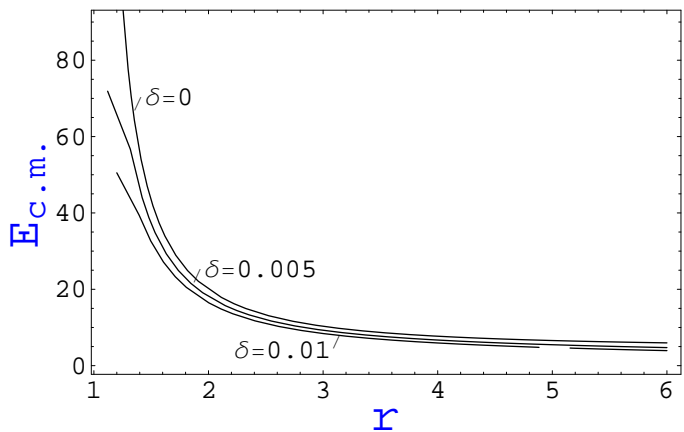}

\caption{\label{ecm} The radial dependence of the center of mass
energy of two infalling particles for the different values of the
parameter $\delta$ in the case of the extreme black hole ($a=M$).
}
\end{figure}

In the Fig. \ref{ecm} the radial dependence of the center mass
energy of two particles for the different values of the
dimensionless parameter $\delta$ has been shown.

From the Fig. \ref{ecm} one can easily see that in the
Ho\v{r}ava--Lifshitz gravity the particle can essentially
accelerate near the horizon but not to arbitrary high energies.
With increasing the parameter $\delta$ the maximal value of the
center of mass energy is decreasing.

{Banados, Silk and West \cite{banados} have shown that the total
energy of two colliding test particles has no upper limit in their
center of mass frame in the neighborhood of an extreme Kerr black
hole, even if these particles were at rest at infinity in the
infinite past. On the contrary we show here that the energy of two
colliding particles in the center of mass frame observed from the
infinity has an upper limit in the Ho\v{r}ava–-Lifshitz gravity. }

\section{\label{conclusion} Conclusion}

We have studied the energetics of the rotating black hole in
Ho\v{r}ava--Lifshitz gravity. Fisrt, we considered the energy
extraction mechanism via the Penrose process and found exact
expression for limit for the change in BH mass (\ref{minenergy})
and concluded that Penrose process can be realized if the
condition $\delta<{2M}/{r_{+}}$ will be satisfied. Since the
parameter $\delta$ is much less than 1, it is easy to conclude
that energy extraction through Penrose process is more realistic
process among the energy extraction mechanisms from BH in
Ho\v{r}ava–-Lifshitz scenario. However, it should be mentioned
that in the early Universe when the module of the cosmological
constant played important role, energy extraction from the
rotating BH could be impossible in Ho\v{r}ava--Lifshitz scenario
due to the positivity of the sign of $E_{\rm min}$. This
limitation for Penrose process does not exist in the standard
theory gravity and appears in the modified theory gravity as
Ho\v{r}ava--Lifshitz one.

In the paper~\cite{banados} authors underlined that rotating black
hole can, in principle, accelerate the particles falling to the
central black hole to arbitrary high energies. Because of some
mechanisms such as astrophysical limitations on the maximal spin,
back-reaction effects, and sensitivity to the initial conditions,
there appears some upper limit for the center of mass energy of
the infalling particles. One of the mechanisms offered in this
paper is appearing due to the Ho\v{r}ava--Lifshitz gravity
correction which prevents particle from the infinite acceleration.

\begin{acknowledgments}

Authors thank the IUCAA for warm hospitality during their stay in
Pune. This research is supported in part by the UzFFR (projects
1-10 and 11-10) and projects FA-F2-F079 and FA-F2-F061 of the
UzAS. This work is partially supported by the ICTP through the
OEA-PRJ-29 project and the TWAS Associateship grant.

\end{acknowledgments}


\begin{thebibliography}{99}


\bibitem{h1}P. Ho\v{r}ava, {JHEP} \textbf {0903},  020 (2009).
%
%
\bibitem{h2}P. Ho\v{r}ava, {Phys. Rev. D} \textbf{ 79}, 084008  (2009).
%

%
\bibitem{CaiCao}
R. G. Cai, L. M. Cao  and N. Ohta, {Phys. Rev. D} \textbf{80},
024003 (2009).
%
\bibitem{Konoplya22}
R. A. Konoplya, {Phys. Lett. B} \textbf{679}, 499 (2009).
%
\bibitem{SChen}
S. Chen and J. Jing, {Phys. Rev. D} \textbf{80}, 024036 (2009).
%
\bibitem{CastilloLar}
A. Castillo and A. Larranaga, Electron. J. Theor. Phys. \textbf{8}
1 (2011).
%
\bibitem{ChenYang}
D. Y. Chen, H. Yang and X. T. Zu, {Phys. Let B} \textbf{681}, 463
(2009).
%
\bibitem{Takashi}

T. Takashi and J. Soda, {Phys. Rev. Lett.} \textbf{102}, 231301
(2009).
%
\bibitem{CalcagniJHEP}
G. Calcagni, {JHEP} \textbf{09},112 (2009). ;
%
\bibitem{Kalyana}
S. Kalyana Rama, {Phys. Rev. D} \textbf{79}, 124031 (2009).
%
\bibitem{Wang}
A. Wang and R. Maartens, {Phys. Rev. D}, \textbf{81}, 024009
(2010).

\bibitem{loboooo}{C. G. Boehmer, L. Hollenstein, F. S. Lobo and S. S. Seahra,
[arXiv:gr-qc/1001.1266].}
%
%
\bibitem{mukohayama}{S. Mukohyama, Class.Quant.Grav. \textbf{27} 223101 (2010).}
%
\bibitem{Saridakis}
E. N. Saridakis, {Eur. Phys. J. C} \textbf{67} 229 (2010).
%
\bibitem{parkkkk}
M. I. Park,     {JCAP} \textbf{1001}, 001 (2010).
%
\bibitem{Mukohyama}
S. Mukohyama, {Physical Review D} \textbf{80}, 064005 (2009).
%
%
\bibitem{loboetal}{T. Harko, Z Kovacs , F.S.N. Lobo, Proc. Roy.
Soc. Lond. A Math. Phys. Eng. Sci. \textbf{467} 1390 (2011). }
%
%
\bibitem{lobo1}
{F.S.N. Lobo, T. Harko and Z. Kova'cs, [arXiv:1001.3517v1
[gr-qc]}.
%
\bibitem{claus}{V. Enolskii, B. Hartmann, V. Kagramanova, J. Kunz, C. Laemmerzahl, P. Sirimachan, [arXiv:1106.2408]}
%
\bibitem{htaa}
{A. Hakimov, B. Turimov, A. Abdujabbarov and B. Ahmedov, {Mod.
Phys. Lett. A}, \textbf{25}, 37 3115 (2010). }
%
\bibitem{aha}{A. Abdujabbarov, B. Ahmedov, and A. Hakimov, {Phys. Rev. D} \textbf{83}, 044053(2011).
}
%
\bibitem{lobo2}
{C.G. B\"{o}hmer and F.S.N. Lobo, Eur. Phys. J. C \textbf{70},
1111 (2010)}. 
%
\bibitem{konoplya11}
{R. A. Konoplya, {Phys. Lett. B} \textbf{679}, 499 (2009). }
%
\bibitem{einstein}{H. L\"{u}, J. Mei, and C.N. Pope, Phys. Rev.
Lett. \textbf{103}, 091302 (2009).}
%
\bibitem{tsoukalas1}{G. Koutsoumbas, E. Papantonopoulos,
P. Pasipoularides, and M. Tsoukalas, Phys. Rev. D \textbf{81},
124014 (2010).}
%
\bibitem{koutsam1}{G. Koutsoumbas, P. Pasipoularides,
Phys. Rev. D  \textbf{82}, 044046 (2010). }

%
\bibitem{dav1}
{M. R. Setare, D. Momeni, Mod. Phys. Lett. A \textbf{26}, 151
(2011).} 
\bibitem{dav2}
{M. R. Setare, D. Momeni, Int. J. Theor. Phys. \textbf{50}, 106
(2011).}
%
%
\bibitem{gwak}{B. Gwak, B.-H. Lee, J. Cosmol. Astropart. Phys. \textbf{09} (2010) 031.}
%
\bibitem{saridakis3}{S. Dutta, E.N. Saridakis, J. Cosmol. Astropart. Phys. {05} (2010) 013;
%
S. Dutta, E.N. Saridakis, J. Cosmol. Astropart. Phys. {01} (2010)
013.}
%
%
\bibitem{ghodsi}{A. Ghodsi, E. Hatefi, {Phys. Rev. D},
\textbf{81}, 044016 (2010). }
%
%
\bibitem{banados}{M. Ba\~{n}ados, J. Silk,  and S.M. West, {Phys.
Rev. Lett.} \textbf{103}, 111102 (2009). }
%
%
\bibitem{berti}{E. Berti, V.  Cardoso, L. Gualtieri, F. Pretorius, and U. Sperhake, {Phys.
Rev. Lett.} \textbf{103}, 239001 (2009). }
%
\bibitem{joshi1}{M. Patil and P. S. Joshi, Phys. Rev. D 82, 104049 (2010);
Phys. Rev. D 83, 064007 (2011).}
%
\bibitem{ghodsi4}{ D. Klemm, V. Moretti and L. Vanzo, {Phys.
Rev. D} \textbf{57}, 6127 (1998). }
%
%
\bibitem{mtw}
C. W. Misner, K.S. Thorne, and J.A. Wheeler, {\it Gravitation}
(San Francisco: Freeman, 1973); K. Prabhu, N. Dadhich, {Phys. Rev.
D} \textbf{81}, 024011  (2010).
%
%
%
\bibitem{penrose69}{R. Penrose, {J. Math. Phys.} \textbf{10}, 38 (1969).}
%
%
%


\end{thebibliography}
\end{document}